\documentclass{article}

\usepackage[english]{babel}

\usepackage[letterpaper,top=2cm,bottom=2cm,left=3cm,right=3cm,marginparwidth=1.75cm]{geometry}

\usepackage{amsmath,amssymb}
\usepackage{url}
\usepackage{amsthm}
\usepackage{graphicx}
\usepackage{tikz}
\usepackage{tkz-berge}
\usepackage{tkz-graph}
\usetikzlibrary{arrows,shapes,positioning}
\usetikzlibrary{decorations.markings}
\usetikzlibrary{calc}
\usepackage{empheq}
\usepackage[ruled,vlined]{algorithm2e}
\usepackage{listings}
\usepackage[tight,footnotesize]{subfigure}
\usepackage[colorlinks=true, allcolors=blue]{hyperref}

\newtheorem{lemma}{Lemma}
\newtheorem{theorem}{Theorem}

\title{Interruption flows for reliability evaluation \\ of power distribution networks}

\author{Fabio Luiz Usberti$^1$, Celso Cavellucci$^1$, Christiano Lyra$^2$\footnote{Corresponding author: chrlyra@unicamp.br} \\ \\ $^1$Institute of Computing, University of Campinas (UNICAMP) \\ $^2$School of Electrical and Computer Engineering, University of Campinas (UNICAMP) \\ }

\begin{document}
\maketitle

\begin{abstract}
Energy
networks should strive for reliability. How can it be assessed, measured, and improved? What are the best trade-offs between investments and their worth? The flow-based framework for the reliability assessment of energy networks 
proposed in this paper addresses these questions with a focus on 
power distribution networks. 
The framework introduces the concept of \textit{iflows}, or interruption flows, which translate the analytical reliability evaluation into solving a series of node balance equations computable in linear time.
The \textit{iflows} permeate the network, providing relevant information to support linear formulations of reliability optimization problems.
Numerical examples showcase the evaluation process obtained through iflows in illustrative distribution networks with distributed generation. A visual representation of the reliability state provides insights into the most critical regions of the network. 
A case study of the optimal allocation of switches in power distribution systems is described. Computational experiments were conducted using a benchmark of distribution networks, having up to 881 nodes. The results confirm the effectiveness of the approach in terms of providing high-quality information and optimal trade-offs to aid reliability decisions for 
energy networks.
\end{abstract}

\section{Introduction} \label{sec:introduction}

The reliability assessment and management of power distribution systems have attracted the attention of governments, utility providers, and the scientific community 
\cite{EscaleraProdanovicCastronuovoRoldanPerez2020}.
There are many reasons for this: approximately $70\%$ of the duration of power supply interruptions originate in distribution networks \cite{billinton}; outages affect the revenues of utilities and shareholders because of the cost of unsupplied energy; continuity standards are established by regulatory agencies, imposing low interruption frequency and duration indices; and mathematical models and efficient methodologies for the planning, operation, and maintenance of smart grids are under development.

The likelihood of consumers being disconnected from their power supply because of outages can be reduced by investing in many areas, e.g., the maintenance of electrical components, location of response teams, signaling devices, allocation of distributed, energy storage devices, and switches.
Computation of the incremental cost of reliability, i.e., the ratio of reliability cost and reliability worth, is an effective tool for determining the economic viability of an investment. 
To ensure that limited capital resources can achieve the best possible outcome requires computationally efficient methodologies that can determine each investment's reliability worth.

\cite{chismant1998} discussed the lack of accurate and efficient methods for evaluating reliability. He proposed a simulation approach for determining the reliability of a utility system. The methodology, despite allowing accurate computation of time-between-failure (TBF) and time-to-repair (TTR) distributions, was computer time intensive.
\cite{Heydt2010}  proposed a Monte Carlo approach to determine the reliability of distribution systems. Computational experiments in which their methodology was compared with an analytical reliability procedure showed that the accuracy of the simulation was relatively high.
\cite{RochaEtal2017} evaluated the effect of distributed generation (DG) on the reliability of distribution networks. The effects of component failures and islanding operations with respect to voltage and frequency variations are considered in their model, which uses simulation to evaluate reliability and assess the dynamics of the islanding process. 
In the same context, \cite{ContiNicolosiRizzo2012} and \cite{AdefaratiBansal2017} proposed probabilistic models to address the stochastic behavior of renewable DG resources and their impact on reliability.

A non-simulated approach was proposed by  \cite{DelgadoContrerasArroyo2018} formulating the reliability assessment of a distribution network as a linear programming optimization problem solving a set of shortest paths in the network. This model was 
 later improved by \cite{TabaresEtal2019}, who proposed a set of linear expressions which can be solved without requiring optimization. 
\cite{LiEtAl2020a} included in their analytical reliability evaluation methodology the assessment of post-fault network reconfiguration. In their experiments, the authors observed that optimal network reconfiguration allows load  transfer  between feeders and significantly improves the reliability of the network. Another facet explored by \cite{Contreras2020} is the switching interruption times, which derives from the isolation of a faulty portion of the network. This is formulated as a mixed-integer programming model, and the computational tests have shown that their model improved the accuracy of the reliability evaluation by considering switching interruptions.

An assumption of the methodologies proposed by \cite{DelgadoContrerasArroyo2018}, \cite{TabaresEtal2019}, \cite{LiEtAl2020a}, and \cite{Contreras2020}  is that all branches of the network are equipped with a switch; this could preclude applications in which the positions of the switches are decision variables.
This requirement was lifted by \cite{LiEtAl2020b} in their optimization-based reliability evaluation methodology. They use the concept of \textit{fictitious flows}, representing the path taken by a fault in the network, calculated by solving an integer linear programming model.

\cite{JuanweiTaoYueXiaohuaBoBaomin2019} assess the reliability of integrated energy systems. They propose an analytical method considering the interdependences between power distribution and gas distribution subsystems, which was shown to be more efficient than simulation approaches.
Comprehensive surveys on computational methodologies for the reliability assessment of distribution systems are provided by \cite{Borges2012} and \cite{Lin2014}.

The demand for fast and accurate reliability evaluation methodologies has been increasing in smart distribution networks \cite{GhianiEtal2018}. These networks, commonly referred to as smart grids, use information and communication technologies, and state estimations of the network to perform automated actions to improve reliability, efficiency, and sustainability of the use of electricity. For example, the reliability can be enhanced with the so-called self-healing of the network, provided by automatic reconfigurations and the capability to perform islanding operations.

\textbf{Contributions}. This research introduces the concept of interruption flows, \textit{iflows} for short, to address the growing demand for a fast analytical reliability evaluation methodology.
The iflows extend the idea of \textit{fictitious flows} brought by \cite{LiEtAl2020b} by expressing relevant reliability states and unfolding new insights into how the interruption circulates throughout a distribution network, indicating its most critical areas. It compares favorably to the state-of-the-art methodologies  in what follows: (\textit{i}) it lifts the requirement of having a switch in every branch of the network, which could preclude some applications; (\textit{ii}) it renders a straightforward evaluation algorithm, with a time complexity that increases linearly with the size of the network; (\textit{iii}) it provides a theoretical foundation from which the iflows can be directly translated to reliability indices and it's corresponding lower and upper bounds; (\textit{iv}) and foremost, it provides linear descriptions of reliability indices that allow strengthening the mathematical formulations of reliability optimization problems. To the best of our knowledge, this is the first reliability evaluation methodology that accomplishes all of the above.

The paper presents with mathematical rigor the equivalence between iflows and the standard reliability evaluation approach \cite{billinton}, and showcases its application in illustrative networks with distributed generation, a feature often present in smart grids. A case study investigates the benefits of iflows in the optimal allocation of switches in radial distribution systems with up to $881$ nodes.
The results illustrate that the methodology can yield optimal trade-offs to support decisions on the best compromises between budgets and reliability. In summary, this paper provides new mathematical and computational concepts that push on the current body of knowledge regarding exact reliability evaluation and optimization methodologies for energy systems.

\section{Terminology and definitions} \label{sec:Terminology}

\subsection{Network representation and terminology} \label{sec:mainConcepts}

A radially operated distribution system can be modeled as a directed tree (arborescence) $G(V,A)$, rooted at the substation (Node $0$) \cite{ahuja93}. Each node $i \in V \setminus \{0\}$ denotes a load point with power load $l_i$ ($kW$), failure rate $\lambda_i$ (failures/year), and number of customers $n_i$. An arc, $(i,j) \in A$, $i,j \in V$, is oriented in the same manner as the power flow, i.e., from the root to the customers. Each node $j \in V \setminus \{ 0 \}$ has a predecessor node $i$, or simply, $i=pred(j)$. The set of arcs that contain switches is denoted by $A_{s}$.

A unique directed path connects the root to every node in the tree. The sequence of nodes representing the directed path connecting two nodes $i$ and $j$ is given by $path(i,j) = \{i, \ldots, j\}$. If no such path exists, then $path(i,j) = \{\emptyset\}$; also, $path(i,i) = \{i\}$. For every pair of nodes $i$ and $j$, if $path(i,j) \neq \{\emptyset\}$, then $j$ is downstream of $i$; otherwise, $j$ is upstream of $i$.
The set of downstream nodes of $i$ is represented by $V_i$. If $V_i = \{i\}$, node $i$ is a leaf. The downstream power load of a node $i$, $\tilde{l}_i$, can be computed without distributed generation by
\begin{align} 
 \displaystyle
 \tilde{l}_i &= \sum_{j \in V_i}{l_j} \label{eq:tildepi}
\end{align}
The necessary  modifications to include distributed generation are discussed in Sec.~\ref{sec:Examples}.

\subsection{Reliability indices} \label{sec:indices}

Regulatory agencies adopt reliability indices to define the minimum levels of reliability, the violation of which can trigger the imposition of fines on utilities. Moreover, reliability indices can also be used to (\textit{i}) identify areas of the network that require additional investment, (\textit{ii}) determine the reliability shifts and trends over time, (\textit{iii}) compare historical values with the values of current network state, and (\textit{iv}) compute the benefit/loss of proposed change to the network \cite{brown2008}.
Several indices are presently used to evaluate  the reliability of a distribution system quantitatively \cite{billinton}, such as the \textit{system average interruption frequency index} (SAIFI), \textit{system average interruption duration index} (SAIDI), and \textit{energy not supplied} (ENS).

Without loss of generality, the ENS (Eq.~\ref{eq:ENS}) is employed in the mathematical developments of the proposed evaluation methodology.

\begin{align} 
ENS = \sum_{i \in V}{\displaystyle l_iu_i} \label{eq:ENS}
\end{align}

Variable $u_i$ represents the duration of interruptions a node $i$ is expected to suffer in a one-year period. A procedure to determine this variable is discussed in Sec.~\ref{sec:Interruption}.

\subsection{Assumptions}

The following assumptions are considered.

\begin{itemize}
	\item The network is radially operated.
	\item All failures are non-transient short circuits that propagate upstream until reaching a switch.
	\item The failure rate is a stochastic parameter, and its value represents the expected amount of failures that should occur in a one-year period.
	\item All switches are automatic sectionalizers with negligible failure rates and operation times. 
\end{itemize}

\subsection{Computation of interruption duration}  \label{sec:Interruption}

This section summarizes the methodology, discussed in depth by \cite{billinton}, to determine the expected duration of power interruptions that follow an outage. 

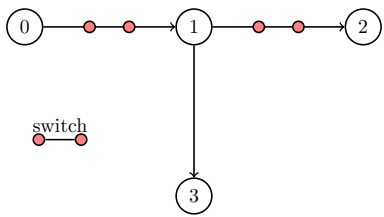
\begin{figure}[hbtp]
\begin{center}
\scalebox{0.75}{
\begin{tikzpicture}
 \SetUpEdge[lw         = 1pt,
            color      = black,
            labelcolor = white]
  \GraphInit[vstyle=Normal] 
  \SetGraphUnit{3}
  \tikzset{VertexStyle/.append  style={fill,thick}}
  \tikzset{mynode/.style=VertexStyle}
  \tikzset{switch/.append style={circle,
         fill=red!50,
         thick,
         inner sep=2pt,
         draw},
        }

  \node[switch] (s1) at (0.25,-2) {};
  \node[switch] (s2) at (1,-2) {};
  \draw[thick,-,black] (s1)--(s2) node [midway,above,black] {switch};

  \node[mynode] (0) at (0,0) {0};
  \node[mynode] (1) at (3,0) {1};
  \node[mynode] (2) at (6,0) {2};
  \node[mynode] (3) at (3,-3) {3};
  \draw[thick,->,black] (0)--(1) node[pos=0.35,switch] {} node[pos=0.65,switch] {};
  \draw[thick,->,black] (1)--(2) node[pos=0.35,switch] {} node[pos=0.65,switch] {};
  \draw[thick,->,black] (1)--(3);
\end{tikzpicture}
}
	\end{center}
	\vspace{-0.5cm}
\caption{Role of a switch.}
\label{fig:example}
\end{figure}

A switch is a  normally closed device that opens when a short circuit flows through it. This event disconnects all downstream load points but prevents power interruption of  upstream loads. Consider the illustrative network shown in Figure~\ref{fig:example}, in which the substation is represented by Node $0$. If a fault occurs at Node $2$, the switch in arc $(1,2)$ opens, interrupting the power supply of Node $2$. However, if the fault occurs at Node $3$, the short circuit propagates up to the substation switch (circuit breaker), causing the interruption of all the load points.

The \textit{expected restoration time} $t_i$ is the expected time required to restore power supply to all the customers affected by a fault in node $i$.
This parameter includes the identification of the failure location, the organization of the maintenance team, the repair of all defective network components, and the reclosure of any switch that opened because of the outage. The IEEE \cite{goldbook} (IEEE Std 493-1997) provides a standard procedure for estimating the restoration time $t_i$ and failure rate $\lambda_i$ for each node $i$ of a distribution network.

The \textit{self-interruption} $\theta_i$ (Eq.~\ref{eq:selftime}) represents the duration of  interruptions that a node $i$ is expected to suffer in a one-year period as a result of local faults (occurring at node $i$). 
	\begin{align} 
		\displaystyle \theta_{i} = \lambda_i t_i \label{eq:selftime} 
	\end{align}

The \textit{downstream interruption} $\tilde{\theta}_{i}$, determined by Eq.~\eqref{eq:downtime},  is a variable representing the time node $i$ is expected to be interrupted because of downstream faults in a one-year period.
	\begin{align}
		\displaystyle \tilde{\theta}_{i} = \theta_{i} + \sum_{(i,j) \in A \setminus A_{s}}{\tilde{\theta}_{j}} \label{eq:downtime}
	\end{align}

As shown by Eq.~\eqref{eq:downtime}, the downstream interruption variables depend on the location of the switches (given by set $A_s$). They can be computed in a bottom-up fashion, starting at the leaves of the network. The downstream interruption of a leaf is simply its self-interruption. In any other node, the downstream interruption comprises the node self-interruption added to the downstream interruptions of every node adjacent to $i$ not isolated by a switch.

The \textit{full interruption} $u_i$ is the duration of interruptions a node $i$ is expected to suffer in a one-year period as a result of all the faults occurring in the network. This variable can be calculated by
\begin{align} 
\displaystyle
u_0 = \tilde{\theta}_{0}, \qquad u_j - u_i =
	\left \{
		\begin{array}{l l}
			0 \qquad & (i,j) \in A \setminus A_{s} \\
			\tilde{\theta}_{j} \qquad & (i,j) \in A_s 
		\end{array}
	\right. \label{eq:fulltime}
\end{align}

Eq.~\eqref{eq:fulltime} shows that the full interruption of the root is equal to its downstream interruption. Moreover, for any arc $(i,j)$ not isolated by a switch, the full interruptions of both its nodes are equal; otherwise, the switch prevents the downstream interruption of $j$ from affecting its predecessor $i$.

To compute the values of full interruptions, a double-sweep procedure can be employed, starting from the bottom-up calculation of the downstream interruptions. Then, a top-down procedure starts from the root, using Eq.~\eqref{eq:fulltime} to determine the full interruption of every node.

\section{Interruption Flows (Iflows)} \label{sec:iflow}

\subsection{Definition}

The interruption flow, or \textit{iflow}, $f_{ij}$, from node $j$ to node $i$ (reverse oriented from the power flow), is the expected duration of interruptions at node $i$ in a one-year period as a result of faults originating at nodes downstream of $j$. Formally, the iflow is defined by
\begin{subequations}
    \begin{empheq}[left={f_{ij}=\empheqlbrace\,}]{align}
      & 0 \qquad \qquad (i,j) \in A_s \label{eq:flow_olda} \\
      & \tilde{\theta}_{j} \qquad \ \ \ \ \ (i,j) \in A \setminus A_{s}         \label{eq:flow_oldb}
    \end{empheq}
\end{subequations}

Eq.~\eqref{eq:flow_olda} shows that an iflow $f_{ij}$ is zero if there is a switch in arc $(i,j)$, which asserts the switch's role in preventing faults downstream of $j$ from affecting node $i$. In the absence of a switch in arc $(i,j)$, Eq.~\eqref{eq:flow_oldb} states that the iflow $f_{ij}$ is equal to the downstream interruption of node $j$, which is consistent with the definition of an iflow.

An alternative identity of an iflow can be obtained by replacing the downstream interruption in Eq.~\eqref{eq:flow_oldb} according to Eq.~\eqref{eq:downtime}.
\begin{align*} 
\displaystyle
f_{ij} =
		\displaystyle \theta_{j} + \sum_{(j,k) \in A \setminus A_{s}}{\tilde{\theta}_{k}}  \qquad \qquad & (i,j) \in A \setminus A_{s} 
\end{align*}

The resulting sum of downstream interruptions can be represented as the sum of iflows, according to Eq.~\eqref{eq:flow_oldb}:
\begin{align*} 
\displaystyle
f_{ij} =
		\displaystyle \theta_{j} + \sum_{(j,k) \in A \setminus A_{s}}{f_{jk}}  \qquad \qquad & (i,j) \in A \setminus A_{s} 
\end{align*}

Because the iflow of an arc containing a switch is zero, the following representation emerges.
\begin{align} 
\displaystyle
f_{ij} =
	\left \{
		\begin{array}{l l}
			0 \\
			\displaystyle \theta_{j} + \sum_{(j,k) \in A}{f_{jk}}   \\
		\end{array}
	\right. \quad
	\left .
		\begin{array}{l l}
			& (i,j) \in A_s \vspace{0.2cm} \\
			& (i,j) \in A \setminus A_{s} \vspace{0.2cm} \\
		\end{array}
	\right.	\qquad	
\label{eq:flow}
\end{align}

\cite{LiEtAl2020b} proposed the idea of \textit{fictitious flows} which represent the path of an interruption from the fault's origin up to the first upstream switch. These flows are defined as binary variables, assuming value equal to one if an interruption flows through the corresponding arc, or zero otherwise. There is a connection between the iflows and \cite{LiEtAl2020b}'s fictitious flows in the sense that they are both equal to zero when a switch is present in an arc, meaning both of them are blocked by the presence of a switch. By capturing the bulk of interruptions in motion, the iflows generalize the fictitious flows, expressing local reliability states and indicating the most critical areas of the network. The iflows also have the advantage of being directly translated into reliability indices, from which lower and upper bounds can be obtained. A theoretical analysis of the iflows is presented in Section~\ref{sec:Eval}.

\subsection{Iflow node balance}

\begin{figure}[hbtp]
	\begin{center}
\scalebox{0.85}{
\begin{tikzpicture}[scale=1,transform shape]
  \Vertex[L=$i$,x=0,y=3]{i}
  \Vertex[L=$j$,x=3,y=3]{j}
  \Vertex[L=$k_1$,x=6,y=5]{k1}
  \Vertex[L=$k_p$,x=6,y=1]{kp}
  \tikzstyle{VertexStyle}=[shape=coordinate]
  \Vertex[x=3,y=4.75]{Fj}
  \Vertex[x=3,y=1.25]{tj}
  \node[] at (3,5) {$F_j$};
  \node[] at (3,1) {$\theta_j$};
  \node[] at (5.25,3) {$\displaystyle \sum_{(j,k) \in A}{f_{jk}}$};
  \node[] at (3.75,3.125) {$\vdots$};
  \tikzstyle{LabelStyle}=[fill=white,sloped,above]
  \tikzstyle{EdgeStyle}=[{<-}]
  \Edge[](i)(j)
  \node[] at (1.5,3.4) {$f_{ij}$};
  \Edge[](j)(k1)
  \Edge[](j)(kp)
  \Edge[](Fj)(j)
  \Edge[](j)(tj)
\end{tikzpicture}
}
	\end{center}
\caption{Iflow node balance.}
\label{fig:balance}
\end{figure}
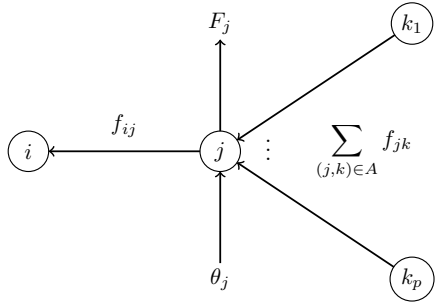

The presence of a switch obstructs the iflow streaming through an arc, and to capture this event within a balance equation, an interruption slack, or \textit{islack}, $F_j$ is defined for each node $j$. Eq.~\eqref{eq:varF} gives the node balance, which is depicted in Fig.~\ref{fig:balance}.
\begin{align}  
\displaystyle f_{ij} + F_j = \theta_j + \sum_{(j,k) \in A}{f_{jk}} \label{eq:varF}
\end{align}

The node balance shows that the values of an iflow $f_{ij}$ and its corresponding islack $F_j$ are complementary. In the presence of a switch, the iflow is zero, while the islack assumes the value that the iflow would take in the absence of the switch.

\subsection{Iflow computation}

The Algorithm~\ref{alg:iflowEval} summarizes the method for computing the iflows. 

\begin{algorithm}[H]
\textbf{Input:} network $G(V,E)$, self-interruption $\theta_u$ for every node $u \in V$, and the root $i$ passed as argument. \\
\textbf{Output:} downstream interruption $\tilde{\theta}_u$ for every node $u \in V$, iflow $f_{uv}$ for every arc $(u,v) \in A$. \\
\Begin
{
    $\tilde{\theta}_i \leftarrow \theta_i$; \\
	\ForAll{$(i,j) \in A$}
	{
	    iflowEval($j$) \\
	    \If{$(i,j) \in A \setminus A_s$}
	    {
	        $\tilde{\theta}_i \leftarrow \tilde{\theta}_i + \tilde{\theta}_j$ \\
	        $f_{ij} \leftarrow \tilde{\theta}_j$
	    }
	    \Else{
	        $f_{ij} \leftarrow 0$
	    }
	}
}
\caption{iflowEval(node $i$)}
\label{alg:iflowEval}
\end{algorithm}

The algorithm implements a recursive depth-first search starting at the root. Once a leaf is reached, the method backtracks carrying the downstream interruption to determine the iflow for the corresponding arc. The time complexity of Algorithm 1 is bounded by $\Theta(|V|+|E|)$, since each node and edge is visited once. This is an improvement over the previous algebraic methods which require solving a set of linear equations \cite{DelgadoContrerasArroyo2018, TabaresEtal2019, LiEtAl2020a, Contreras2020}.

Figure~\ref{fig:code-execution} shows the execution steps of Algorithm~1 applied to the illustrative network depicted in Figure~\ref{fig:example}. It is assumed that the self-interruptions of Nodes 0, 1, 2, and 3 ($\theta_0, \theta_1, \theta_2, \theta_3$, respectively) are known. Step 1 starts by calling the method at the root, Node 0. The execution then recursively dives into the network until a leaf is reached. At Step 7 it arrives at Node 2, a leaf. From there, it backtracks from Arc $(1,2)$, setting the value of the corresponding iflow to zero ($f_{12} = 0$), since the iflow was blocked by a switch. Similarly, at Step 12 the execution reaches Node 3, another leaf. It then backtracks from Arc $(1,3)$, setting the value of the corresponding iflow in accordance with the value of the downstream interruption ($f_{13} = \tilde{\theta}_3$). Finally, the execution backtracks one last time from Arc $(0,1)$, setting the iflow to zero ($f_{01} = 0$). 

\begin{figure}
    \centering
    \begin{lstlisting}[mathescape]
    iflowEval(0)
        $\tilde{\theta}_0 \leftarrow \theta_0$
        // dive to Arc (0,1)
        iflowEval(1)
            $\tilde{\theta}_1 \leftarrow \theta_1$
            // dive to Arc (1,2)
            iflowEval(2)
                $\tilde{\theta}_2 \leftarrow \theta_2$
            // backtrack from Arc (1,2)
            $f_{12} \leftarrow 0$
            // dive to Arc (1,3)
            iflowEval(3)
                $\tilde{\theta}_3 \leftarrow \theta_3$
            // backtrack from Arc (1,3)
            $\tilde{\theta}_1 \leftarrow \tilde{\theta_1} + \tilde{\theta_3}$
            $f_{13} \leftarrow \tilde{\theta}_3$
        // backtrack from Arc (0,1)
        $f_{01} \leftarrow 0$
\end{lstlisting}
\vspace{-1cm}
    \caption{Execution of Algorithm 1 to the network depicted in Figure~\ref{fig:example}.}
    \label{fig:code-execution}
\end{figure}

The following section describes the extraction of reliability measures for a network once the iflows have been obtained. 

\section{Flow-based reliability evaluation} \label{sec:Eval}

This section revisits the ENS reliability index under the perspective of iflows, which allows the inference of lower and upper bounds for the ENS.
The evaluation of other reliability indices with iflows is briefly discussed later in this section.

\subsection{Formulating ENS with iflows} \label{ssec:ENSiflow}

Lemma~\ref{lemma1} shows that the interruption times $u$ can be expressed in terms of the islacks.

\begin{lemma} \label{lemma1}
	The full interruption $u_j$ of a node $j$ can be expressed as
	\begin{subequations}
	\begin{align} 
		& \displaystyle u_0 = F_0 \label{eq:ttimeFrecA} \\
		& \displaystyle u_j - u_i = F_j \qquad (i,j) \in A \label{eq:ttimeFrecB}
	\end{align}
	\end{subequations}
\end{lemma}

\begin{proof}
    Proof of Eq.~\eqref{eq:ttimeFrecA} (root):
    \begin{align}
	    \displaystyle
	    F_0 &\stackrel{\mbox{\tiny Eq.~\ref{eq:varF}}}{=} \theta_0 + \sum_{(0,i) \in A}{f_{0i}} \stackrel{\mbox{\tiny Eq.~\ref{eq:flow_oldb}}}{=} \theta_0 + \sum_{(0,i) \in A \setminus A_{s}}{\tilde{\theta}_{i}} \stackrel{\mbox{\tiny Eq.~\ref{eq:downtime}}}{=} \tilde{\theta}_{0} \stackrel{\mbox{\tiny Eq.~\ref{eq:fulltime}}}{=} u_0 \nonumber
    \end{align}
    Proof of Eq.~\eqref{eq:ttimeFrecB} when $(i,j) \in A \setminus A_{s}$:
    \begin{align}  
	    \displaystyle
	    F_j &\stackrel{\mbox{\tiny Eq.~\ref{eq:varF}}}{=} \displaystyle \theta_j + \sum_{(j,k) \in A}{f_{jk}} - f_{ij} \stackrel{\mbox{\tiny Eq.~\ref{eq:flow_oldb}}}{=} \theta_j +\sum_{(j,k) \in A \setminus A_{s}}{\tilde{\theta}_{k}} - \tilde{\theta}_{j} \nonumber \\
	    &\stackrel{\mbox{\tiny Eq.~\ref{eq:downtime}}}{=} \tilde{\theta}_{j} - \tilde{\theta}_{j} = 0 \stackrel{\mbox{\tiny Eq.~\ref{eq:fulltime}}}{=} u_j - u_i \nonumber	
    \end{align}
    Proof of Eq.~\eqref{eq:ttimeFrecB} when $(i,j) \in A_{s}$:
    \begin{align}  
	    \displaystyle
        F_j &\stackrel{\mbox{\tiny Eq.~\ref{eq:varF}}}{=} \displaystyle \theta_j + \sum_{(j,k) \in A}{f_{jk}} - f_{ij} \nonumber  \\
        &\stackrel{\mbox{\tiny Eq.~\ref{eq:flow_oldb}}}{=} \theta_j +\sum_{(j,k) \in A \setminus A_{s}}{\tilde{\theta}_{k}} \stackrel{\mbox{\tiny Eq.~\ref{eq:downtime}}}{=} \tilde{\theta}_{j} \stackrel{\mbox{\tiny Eq.~\ref{eq:fulltime}}}{=} u_j - u_i \nonumber
    \end{align}
\end{proof}

Eq.~\eqref{eq:ttimeFrecB} is a recursive expression that links full interruptions and islacks. Lemma~\ref{lemma2} removes this recursion to attain a more straightforward expression.

\begin{lemma} \label{lemma2}
	The full interruption $u_j$ is the sum of the islacks on the path from the root to node $j$:
	\begin{align} \label{eq:ttimeF2}
		\displaystyle u_j = \sum_{i \in path(0,j)}{F_i} \qquad \qquad j \in V
	\end{align}
\end{lemma}

\begin{proof}
	The proof is through induction on the path from the root to node $j$. First, the base case, where node $j$ is the root ($j=0$), is proven.
 	\begin{align}
 	    u_0 &\stackrel{\mbox{\tiny Lem.~\ref{lemma1}}}{=} F_0 = \sum_{j \in path(0,0)}{F_j} \nonumber
	\end{align}
	As the induction hyphothesis, it is assumed that Eq.~\eqref{eq:ttimeF2} holds for any node in the path from the root to node $a$. Now, we prove that Eq.~\eqref{eq:ttimeF2} also holds for node $b$, such that $(a,b) \in A$, starting with the expression given by Eq.~\eqref{eq:ttimeFrecB}:
 	\begin{align}
     	\displaystyle
 	    u_b - u_a = F_b \quad &\Rightarrow \quad u_b - \sum_{i \in path(0,a)}{F_i} = F_b \nonumber \quad \nonumber \\ &\Rightarrow \quad u_b = \sum_{i \in path(0,b)}{F_i} \nonumber
	\end{align}
\end{proof}

Theorem~\ref{theo1} presents the core of the reliability evaluation methodology. It shows the ENS expressed in terms of the iflows. 

\begin{theorem} \label{theo1}
The ENS can be expressed in terms of the iflows, as follows.
 	\begin{align} \label{eq:ENS2}
		\displaystyle ENS = \sum_{(i,j) \in A}{(\tilde{l}_i-\tilde{l}_j)f_{ij}} + \sum_{i \in V}{\tilde{l}_i\theta_{i}}
	\end{align}
\end{theorem}
\begin{proof}
	\begin{subequations} 
	\begin{align}
        \displaystyle \sum_{i \in V}{l_iu_i} &= \sum_{i \in V}{\left( l_i\sum_{j \in path(0,i)}{F_j} \right)} \label{eq:ens_a}  \\
		&= \sum_{j \in V}{\left( F_j\sum_{i \in V_j}{l_i} \right)} \label{eq:ens_b} \\
		&= \sum_{i \in V}{\tilde{l}_iF_i} \label{eq:ens_c} \\
		&= \sum_{i \in V}{\left( \tilde{l}_i\left(\theta_i + \sum_{(i,j) \in A}{f_{ij}} \right) \right)} - \sum_{(i,j) \in A}{\tilde{l}_jf_{ij}} \label{eq:ens_d}	 \\
		&= \sum_{(i,j) \in A}{(\tilde{l}_i-\tilde{l}_j)f_{ij}} + \sum_{i \in V}{\tilde{l}_i\theta_{i}} \label{eq:ens_e}
	\end{align}
	\end{subequations}
\end{proof}

For the proof of Theorem~\ref{theo1}, first the definition of ENS is utilized, as stated in Eq.~\eqref{eq:ENS}. Eq.~\eqref{eq:ens_a} comes from the equivalence between full interruptions and islacks, given by Eq.~\eqref{eq:ttimeF2}. Eq.~\eqref{eq:ens_b} is derived by the fact that node $j$ belongs to the path from the root to node $i$ ($j \in path(0,i)$) if and only if node $i$ is downstream of node $j$ ($i \in V_j$).  Eq.~\eqref{eq:ens_c} comes from the definition of the downstream power load (Eq.~\ref{eq:tildepi}). By using the iflow node balance given by Eq.~\eqref{eq:varF}, Eq.~\eqref{eq:ens_d} is obtained. Finally, Eq.~\eqref{eq:ens_e} can be achieved through the algebraic rearrangement of terms. 

\subsection{Lower and upper bounds} 

\label{ssec:lbub}

It is worth mentioning that the term $\displaystyle \sum_{i \in V}{\tilde{l}_i\theta_{i}}$ (Eq.~\ref{eq:ENS2}) is a constant, independent of the iflows, and represents a lower bound for the ENS.
Thus, the minimum ENS that can be expected in a network is obtained by taking the self-interruption of each node multiplied by its downstream load. This observation is extended in Lemma~\ref{lem:Eminmax} by showing that the ENS value has a feasible interval, expressed by a lower and an upper bound. The relative reliability state of a network can be established in terms of the distance from these bounds.

\begin{lemma} \label{lem:Eminmax}
A lower bound $E_{lb}$ and upper bound $E_{ub}$ for the ENS are described by the following inequalities.
 	\begin{align} 
		\displaystyle \sum_{i \in V}{\tilde{l}_i\theta_{i}} = E_{lb} \leqslant ENS \leqslant E_{ub} = \tilde{l}_0\sum_{i \in V}{\theta_{i}} \label{eq:Eminmax}
	\end{align}
\end{lemma}
\begin{proof}
As discussed previously, the ENS lower bound $E_{lb}$ is trivially inferred as the constant in Eq.~\eqref{eq:ENS2}. The scenario in which $ENS = E_{lb}$ implies that all arcs contain a switch ($A_{s} = A$), and thus, $f_{ij} = 0$ for every arc $(i,j)$.

With respect to the ENS upper bound $E_{ub}$, an expression can be derived by assuming that the value of the iflow of every arc $(i,j)$ is maximum, which occurs when there is no switch in the network ($A_{s} = \emptyset$). Under this assumption, the maximum iflow of an arc $(i,j)$, $f_{ij}^{max}$, is given by
 	\begin{align} \label{eq:maxflow}
		f^{max}_{ij} \stackrel{\mbox{\tiny Eq.~\ref{eq:flow}}}{=} \tilde{\theta}_{j} \stackrel{\mbox{\tiny Eq.~\ref{eq:downtime}}}{=} \theta_j + \sum_{(i,j) \in A \setminus A_{s}}{\tilde{\theta}_{j}} \stackrel{A_{s} = \emptyset}{=} \sum_{k \in V_j}{\theta_{k}} \qquad (i,j) \in A
	\end{align}

The maximum iflow of an arc $(i,j)$ is thus the sum of self-interruptions from all nodes downstream from $j$. Using this knowledge, the $E_{ub}$ can be derived

\begin{subequations}
 	\begin{align} 
		\displaystyle &E_{ub} = \sum_{(i,j) \in A}{(\tilde{l}_i-\tilde{l}_j)f^{max}_{ij}} + \sum_{i \in V}{\tilde{l}_i\theta_{i}} \label{eq:eub_a}  \\
		&= \sum_{(i,j) \in A}{\left( \tilde{l}_i\sum_{k \in V_j}{\theta_k} \right)} - \sum_{(i,j) \in A}{\left( \tilde{l}_j\sum_{k \in V_j}{\theta_k} \right)} + \sum_{i \in V}{\tilde{l}_i\theta_{i}} \label{eq:eub_b}  \\
		&= \sum_{i \in V}{\left( \tilde{l}_i \left( \sum_{j \in V_i}{\theta_j} - \theta_i \right) \right)} - \sum_{i \in V \setminus {0}}{\left( \tilde{l}_i\sum_{j \in V_i}{\theta_j} \right)} + \sum_{i \in V}{\tilde{l}_i\theta_{i}} \label{eq:eub_c} \\
		&= \tilde{l}_0 \left( \sum_{i \in V}{\theta_i} - \theta_0 \right) + \sum_{i \in V \setminus {0}}{\tilde{l}_i \left( \sum_{j \in V_i}{\theta_j} - \theta_i \right) } \nonumber \\ 
		& \ \ \ - \sum_{i \in V \setminus {0}}{\left( \tilde{l}_i\sum_{j \in V_i}{\theta_j} \right)} + \sum_{i \in V}{\tilde{l}_i\theta_{i}} \label{eq:eub_d}  \\		
		&= \tilde{l}_0\sum_{i \in V}{\theta_i} - \tilde{l}_0\theta_0 - \sum_{i \in V \setminus {0}}{\tilde{l}_i\theta_i} + \sum_{i \in V}{\tilde{l}_i\theta_{i}}  = \tilde{l}_0\sum_{i \in V}{\theta_i} \label{eq:eub_e}
	\end{align}
\end{subequations}
\end{proof}

Eq.~\eqref{eq:eub_a} is derived by entering the maximum iflows in Eq.~\eqref{eq:ENS2}. Eq.~\eqref{eq:eub_b} follows by using the maximum iflow expression given in Eq.~\eqref{eq:maxflow}. By rewriting the summations over the arcs in the summations over the nodes, Eq.~\eqref{eq:eub_c} is obtained. To obtain Eq.~\eqref{eq:eub_d}, the root is detached from the summation over the nodes. Finally, Eq.~\eqref{eq:eub_e} can be achieved through the algebraic rearrangement of terms. 

The $E_{ub}$ expression provides the worst possible reliability for any network, which in terms of ENS represents the sum of all loads multiplied by the sum of all self-interruptions. Conversely, the $E_{lb}$ value gives the reliability that a network can attain if every interruption is contained in the best possible manner. This can be achieved by locating a switch in every arc, a scenario considered by \cite{TabaresEtal2019} in their analytical reliability evaluation methodology. The $E_{lb}$ expression encapsulates the solution of the linear system of equations that appears in \cite{TabaresEtal2019} methodology.

The iflows do not have the prior requirement of a switch in every arc to compute the network reliability.  This attribute is an advantage  to model problems in which the switches locations are decision variables, as will be shown in Sec.~\ref{sec:Optimization}.

\subsection{Other reliability indices} \label{ssec:otherIndices}

The formal analysis described in this section can also be extended to other reliability indices. In fact, any index tied to interruption times $u$, such as SAIDI, can be expressed in terms of iflows by replacing $u$ with the islacks (Eq.~\ref{eq:ENS}) and then using the iflow node balance (Eq.~\ref{eq:varF}). Also, lower and upper bounds can be inferred by the same analysis. 
With respect to indices that rely solely on interruption frequencies, such as SAIFI, it should be noticed that the frequency and duration of interruptions are correlated by the restoration time $t$ (Eq.~\ref{eq:selftime}). By using this correlation, the iflow magnitude can be translated from \textit{units of time} to \textit{units of frequency}, allowing the computation of indices such as SAIFI.

\section{Illustrative examples with iflows} \label{sec:Examples}

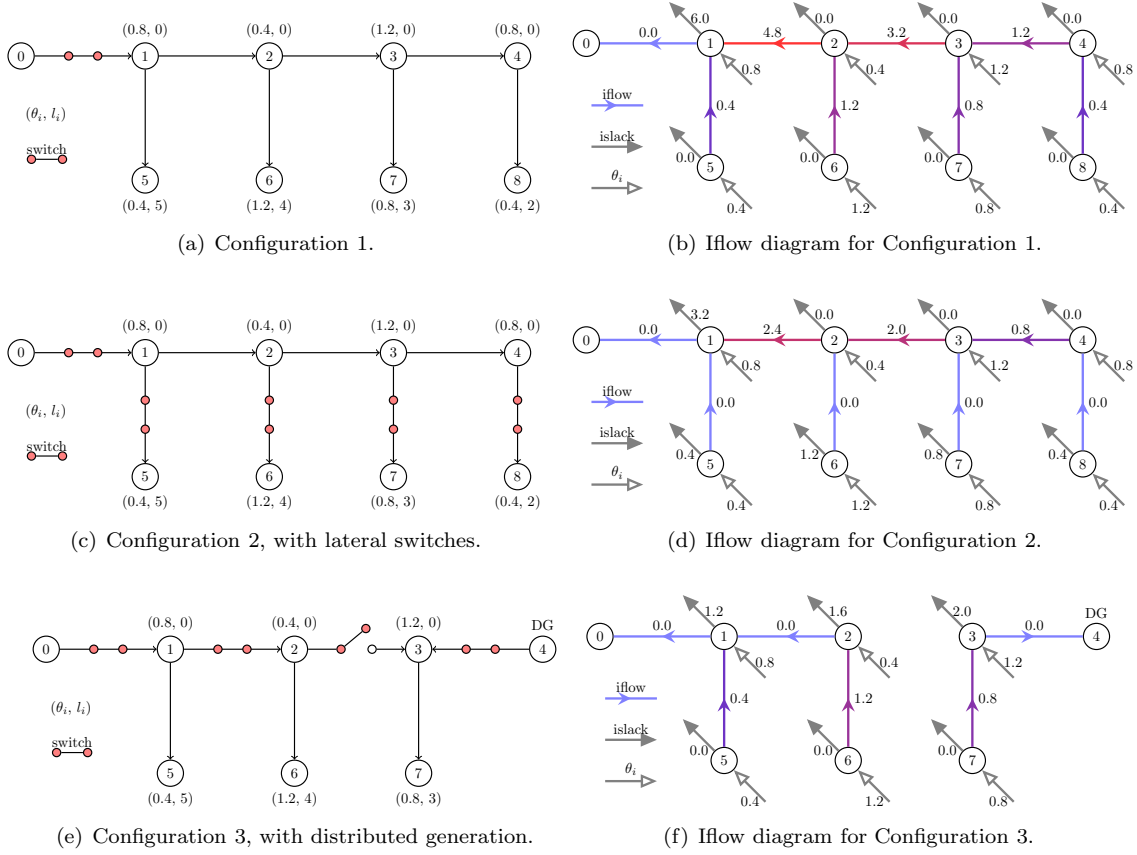
\begin{figure*}[hbtp]
\begin{center}
  \subfigure[{Configuration 1.} \label{fig:billintonCaseE1a}]
      {\scalebox{0.55}{
\begin{tikzpicture}
 \SetUpEdge[lw         = 1pt,
            color      = black,
            labelcolor = white]
  \GraphInit[vstyle=Normal] 
  \SetGraphUnit{3}
  \tikzset{VertexStyle/.append  style={fill,thick}}
  \tikzset{mynode/.style=VertexStyle}
  \tikzset{switch/.append style={circle,
         fill=red!50,
         thick,
         inner sep=2pt,
         draw},
        }

  \node[label=below:{($\theta_i$, $l_i$)}] at (0.6,-1) {};
  \node[switch] (s1) at (0.25,-2.5) {};
  \node[switch] (s2) at (1,-2.5) {};
  \draw[thick,-,black] (s1)--(s2) node [midway,above,black] {switch};

  \node[mynode] (0) at (0,0) {0};
  \node[mynode,label={(0.8, 0)}] (1) at (3,0) {1};
  \node[mynode,label={(0.4, 0)}] (2) at (6,0) {2};
  \node[mynode,label={(1.2, 0)}] (3) at (9,0) {3};
  \node[mynode,label={(0.8, 0)}] (4) at (12,0) {4};
  \node[mynode,label=below:{(0.4, 5)}] (5) at (3,-3) {5};
  \node[mynode,label=below:{(1.2, 4)}] (6) at (6,-3) {6};
  \node[mynode,label=below:{(0.8, 3)}] (7) at (9,-3) {7};
  \node[mynode,label=below:{(0.4, 2)}] (8) at (12,-3) {8};
  \draw[thick,->,black] (0)--(1) node[pos=0.35,switch] {} node[pos=0.65,switch] {};
  \draw[thick,->,black] (1)--(2);
  \draw[thick,->,black] (2)--(3);
  \draw[thick,->,black] (3)--(4);
  \draw[thick,->,black] (1)--(5);
  \draw[thick,->,black] (2)--(6);
  \draw[thick,->,black] (3)--(7);
  \draw[thick,->,black] (4)--(8);
\end{tikzpicture}
}} 
  \subfigure[{Iflow diagram for Configuration 1.} \label{fig:billintonCaseE1b}]
      {\scalebox{0.55}{
\begin{tikzpicture}
 \SetUpEdge[lw         = 1pt,
            color      = black,
            labelcolor = white]
  \GraphInit[vstyle=Normal] 

  \node[] (s1) at (0,-1.5) {};
  \node[] (s2) at (1.5,-1.5) {};
  \draw[ultra thick,directed1,blue!50] (s1)--(s2) node [midway,above,black] {iflow};
  \node[] (s3) at (0,-2.5) {};
  \node[] (s4) at (1.5,-2.5) {};
  \draw[ultra thick,-triangle 45,black!50] (s3)--(s4) node [midway,above,black] {islack};
  \node[] (s5) at (0,-3.5) {};
  \node[] (s6) at (1.5,-3.5) {};
  \draw[ultra thick,-open triangle 45,black!50] (s5)--(s6) node [midway,above,black] {$\theta_i$};

  \SetGraphUnit{3}
  \tikzset{VertexStyle/.append  style={fill,thick}}
  \Vertex[x=0 ,y=0]{0}
  \Vertex[x=3 ,y=0]{1}
  \Vertex[x=6 ,y=0]{2}
  \Vertex[x=9 ,y=0]{3}
  \Vertex[x=12 ,y=0]{4}
  \Vertex[x=3 ,y=-3]{5}
  \Vertex[x=6 ,y=-3]{6}
  \Vertex[x=9 ,y=-3]{7}
  \Vertex[x=12 ,y=-3]{8}
  \tikzset{EdgeStyle/.style={->}}
  \draw[ultra thick,directed1,red!0!blue!50] (1)--(0) node [midway,above,black] {$0.0$};
  \draw[ultra thick,directed1,red!100!blue!80] (2)--(1) node [midway,above,black] {$4.8$};
  \draw[ultra thick,directed1,red!82!blue!80] (3)--(2) node [midway,above,black] {$3.2$};
  \draw[ultra thick,directed1,red!50!blue!80] (4)--(3) node [midway,above,black] {$1.2$};
  \draw[ultra thick,directed1,red!29!blue!80] (5)--(1) node [midway,right,black] {$0.4$};
  \draw[ultra thick,directed1,red!50!blue!80] (6)--(2) node [midway,right,black] {$1.2$};
  \draw[ultra thick,directed1,red!41!blue!80] (7)--(3) node [midway,right,black] {$0.8$};
  \draw[ultra thick,directed1,red!29!blue!80] (8)--(4) node [midway,right,black] {$0.4$};
  
  \draw [ultra thick,black!50!white,-triangle 45] (1) -- +(-1,1) node [midway,right,black] {$6.0$};
  \draw [ultra thick,black!50!white,-open triangle 45] ($(1) +(1,-1)$) -- (1) node [midway,right,black] {$0.8$};
  \draw [ultra thick,black!50!white,-triangle 45] (2) -- +(-1,1) node [midway,right,black] {$0.0$};
  \draw [ultra thick,black!50!white,-open triangle 45] ($(2) +(1,-1)$) -- (2) node [midway,right,black] {$0.4$};
  \draw [ultra thick,black!50!white,-triangle 45] (3) -- +(-1,1) node [midway,right,black] {$0.0$};
  \draw [ultra thick,black!50!white,-open triangle 45] ($(3) +(1,-1)$) -- (3) node [midway,right,black] {$1.2$};
  \draw [ultra thick,black!50!white,-triangle 45] (4) -- +(-1,1) node [midway,right,black] {$0.0$};
  \draw [ultra thick,black!50!white,-open triangle 45] ($(4) +(1,-1)$) -- (4) node [midway,right,black] {$0.8$};
  \draw [ultra thick,black!50!white,-triangle 45] (5) -- +(-1,1) node [at start,left,black] {$0.0$};
  \draw [ultra thick,black!50!white,-open triangle 45] ($(5) +(1,-1)$) -- (5) node [at start,left,black] {$0.4$};
  \draw [ultra thick,black!50!white,-triangle 45] (6) -- +(-1,1) node [at start,left,black] {$0.0$};
  \draw [ultra thick,black!50!white,-open triangle 45] ($(6) +(1,-1)$) -- (6) node [at start,left,black] {$1.2$};
  \draw [ultra thick,black!50!white,-triangle 45] (7) -- +(-1,1) node [at start,left,black] {$0.0$};
  \draw [ultra thick,black!50!white,-open triangle 45] ($(7) +(1,-1)$) -- (7) node [at start,left,black] {$0.8$};
  \draw [ultra thick,black!50!white,-triangle 45] (8) -- +(-1,1) node [at start,left,black] {$0.0$};
  \draw [ultra thick,black!50!white,-open triangle 45] ($(8) +(1,-1)$) -- (8) node [at start,left,black] {$0.4$};
\end{tikzpicture}
}} 
\end{center}
\begin{center}
  \subfigure[{Configuration 2, with lateral switches.} \label{fig:billintonCaseE2a}]
      {\scalebox{0.55}{
\begin{tikzpicture}
 \SetUpEdge[lw         = 1pt,
            color      = black,
            labelcolor = white]
  \GraphInit[vstyle=Normal] 
  \SetGraphUnit{3}
  \tikzset{VertexStyle/.append  style={fill,thick}}
  \tikzset{mynode/.style=VertexStyle}
  \tikzset{switch/.append style={circle,
         fill=red!50,
         thick,
         inner sep=2pt,
         draw},
        }
        
  \node[mynode] (0) at (0,0) {0};
  \node[mynode,label={(0.8, 0)}] (1) at (3,0) {1};
  \node[mynode,label={(0.4, 0)}] (2) at (6,0) {2};
  \node[mynode,label={(1.2, 0)}] (3) at (9,0) {3};
  \node[mynode,label={(0.8, 0)}] (4) at (12,0) {4};
  \node[mynode,label=below:{(0.4, 5)}] (5) at (3,-3) {5};
  \node[mynode,label=below:{(1.2, 4)}] (6) at (6,-3) {6};
  \node[mynode,label=below:{(0.8, 3)}] (7) at (9,-3) {7};
  \node[mynode,label=below:{(0.4, 2)}] (8) at (12,-3) {8};
  
  \node[label=below:{($\theta_i$, $l_i$)}] at (0.6,-1) {};
  \node[switch] (s1) at (0.25,-2.5) {};
  \node[switch] (s2) at (1,-2.5) {};
  \draw[thick,-,black] (s1)--(s2) node [midway,above,black] {switch};
  
  \draw[thick,->,black] (0)--(1) node[pos=0.35,switch] {} node[pos=0.65,switch] {};
  \draw[thick,->,black] (1)--(2);
  \draw[thick,->,black] (2)--(3);
  \draw[thick,->,black] (3)--(4);
  \draw[thick,->,black] (1)--(5) node[pos=0.35,switch] {} node[pos=0.65,switch] {};
  \draw[thick,->,black] (2)--(6) node[pos=0.35,switch] {} node[pos=0.65,switch] {};
  \draw[thick,->,black] (3)--(7) node[pos=0.35,switch] {} node[pos=0.65,switch] {};
  \draw[thick,->,black] (4)--(8) node[pos=0.35,switch] {} node[pos=0.65,switch] {};
\end{tikzpicture}
}} 
  \subfigure[{Iflow diagram for Configuration 2.} \label{fig:billintonCaseE2b}]
      {\scalebox{0.55}{
\begin{tikzpicture}
 \SetUpEdge[lw         = 1pt,
            color      = black,
            labelcolor = white]
  \GraphInit[vstyle=Normal] 
  \SetGraphUnit{3}
  
  \node[] (s1) at (0,-1.5) {};
  \node[] (s2) at (1.5,-1.5) {};
  \draw[ultra thick,directed1,blue!50] (s1)--(s2) node [midway,above,black] {iflow};
  \node[] (s3) at (0,-2.5) {};
  \node[] (s4) at (1.5,-2.5) {};
  \draw[ultra thick,-triangle 45,black!50] (s3)--(s4) node [midway,above,black] {islack};
  \node[] (s5) at (0,-3.5) {};
  \node[] (s6) at (1.5,-3.5) {};
  \draw[ultra thick,-open triangle 45,black!50] (s5)--(s6) node [midway,above,black] {$\theta_i$};

  \tikzset{VertexStyle/.append  style={fill,thick}}
  \Vertex[x=0 ,y=0]{0}
  \Vertex[x=3 ,y=0]{1}
  \Vertex[x=6 ,y=0]{2}
  \Vertex[x=9 ,y=0]{3}
  \Vertex[x=12 ,y=0]{4}
  \Vertex[x=3 ,y=-3]{5}
  \Vertex[x=6 ,y=-3]{6}
  \Vertex[x=9 ,y=-3]{7}
  \Vertex[x=12 ,y=-3]{8}

  \tikzset{EdgeStyle/.style={->}}
  \draw[ultra thick,directed1,red!0!blue!50] (1)--(0) node [midway,above,black] {$0.0$};
  \draw[ultra thick,directed1,red!71!blue!80] (2)--(1) node [midway,above,black] {$2.4$};
  \draw[ultra thick,directed1,red!65!blue!80] (3)--(2) node [midway,above,black] {$2.0$};
  \draw[ultra thick,directed1,red!40!blue!80] (4)--(3) node [midway,above,black] {$0.8$};
  \draw[ultra thick,directed1,red!0!blue!50] (5)--(1) node [midway,right,black] {$0.0$};
  \draw[ultra thick,directed1,red!0!blue!50] (6)--(2) node [midway,right,black] {$0.0$};
  \draw[ultra thick,directed1,red!0!blue!50] (7)--(3) node [midway,right,black] {$0.0$};
  \draw[ultra thick,directed1,red!0!blue!50] (8)--(4) node [midway,right,black] {$0.0$};

  \draw [ultra thick,black!50!white,-triangle 45] (1) -- +(-1,1) node [midway,right,black] {$3.2$};
  \draw [ultra thick,black!50!white,-open triangle 45] ($(1) +(1,-1)$) -- (1) node [midway,right,black] {$0.8$};
  \draw [ultra thick,black!50!white,-triangle 45] (2) -- +(-1,1) node [midway,right,black] {$0.0$};
  \draw [ultra thick,black!50!white,-open triangle 45] ($(2) +(1,-1)$) -- (2) node [midway,right,black] {$0.4$};
  \draw [ultra thick,black!50!white,-triangle 45] (3) -- +(-1,1) node [midway,right,black] {$0.0$};
  \draw [ultra thick,black!50!white,-open triangle 45] ($(3) +(1,-1)$) -- (3) node [midway,right,black] {$1.2$};
  \draw [ultra thick,black!50!white,-triangle 45] (4) -- +(-1,1) node [midway,right,black] {$0.0$};
  \draw [ultra thick,black!50!white,-open triangle 45] ($(4) +(1,-1)$) -- (4) node [midway,right,black] {$0.8$};
  \draw [ultra thick,black!50!white,-triangle 45] (5) -- +(-1,1) node [at start,left,black] {$0.4$};
  \draw [ultra thick,black!50!white,-open triangle 45] ($(5) +(1,-1)$) -- (5) node [at start,left,black] {$0.4$};
  \draw [ultra thick,black!50!white,-triangle 45] (6) -- +(-1,1) node [at start,left,black] {$1.2$};
  \draw [ultra thick,black!50!white,-open triangle 45] ($(6) +(1,-1)$) -- (6) node [at start,left,black] {$1.2$};
  \draw [ultra thick,black!50!white,-triangle 45] (7) -- +(-1,1) node [at start,left,black] {$0.8$};
  \draw [ultra thick,black!50!white,-open triangle 45] ($(7) +(1,-1)$) -- (7) node [at start,left,black] {$0.8$};
  \draw [ultra thick,black!50!white,-triangle 45] (8) -- +(-1,1) node [at start,left,black] {$0.4$};
  \draw [ultra thick,black!50!white,-open triangle 45] ($(8) +(1,-1)$) -- (8) node [at start,left,black] {$0.4$};

\end{tikzpicture}
}} 
\end{center}
\begin{center}
  \subfigure[{Configuration 3, with distributed generation.} \label{fig:billintonCaseE3a}]
      {\scalebox{0.55}{
\begin{tikzpicture}
 \SetUpEdge[lw         = 1pt,
            color      = black,
            labelcolor = white]
  \GraphInit[vstyle=Normal] 
  \SetGraphUnit{3}
  \tikzset{VertexStyle/.append  style={fill,thick}}
  \tikzset{mynode/.style=VertexStyle}
  \tikzset{switch/.append style={circle,
         fill=red!50,
         thick,
         inner sep=2pt,
         draw},
        }
    \tikzset{openswitch/.append style={circle,
         fill=white,
         thick,
         inner sep=2pt,
         draw},
        }
        
  \node[mynode] (0) at (0,0) {0};
  \node[mynode,label={(0.8, 0)}] (1) at (3,0) {1};
  \node[mynode,label={(0.4, 0)}] (2) at (6,0) {2};
  \node[mynode,label={(1.2, 0)}] (3) at (9,0) {3};
  \node[mynode,label={DG}] (4) at (12,0) {4};
  \node[mynode,label=below:{(0.4, 5)}] (5) at (3,-3) {5};
  \node[mynode,label=below:{(1.2, 4)}] (6) at (6,-3) {6};
  \node[mynode,label=below:{(0.8, 3)}] (7) at (9,-3) {7};

  \node[label=below:{($\theta_i$, $l_i$)}] at (0.6,-1) {};
  \node[switch] (s1) at (0.25,-2.5) {};
  \node[switch] (s2) at (1,-2.5) {};
  \draw[thick,-,black] (s1)--(s2) node [midway,above,black] {switch};
  
  \draw[thick,->,black] (0)--(1) node[pos=0.35,switch] {} node[pos=0.65,switch] {};
  \draw[thick,->,black] (1)--(2) node[pos=0.35,switch] {} node[pos=0.65,switch] {};
  \draw[thick,->,black] (4)--(3) node[pos=0.35,switch] {} node[pos=0.65,switch] {};
  \draw[thick,->,black] (1)--(5);
  \draw[thick,->,black] (2)--(6);
  \draw[thick,->,black] (3)--(7);
  
  \node[switch] (s3) at (6+3*0.375,0) {};
  \node[openswitch] (s4) at (6+3*0.625,0) {};
  \node[switch] (s5) at (6+3*0.575,0.5) {};
  \draw[thick,-,black] (s3)--(s5);
  \draw[thick,-,black] (2)--(s3);
  \draw[thick,->,black] (s4)--(3);
\end{tikzpicture}
}} 
  \subfigure[{Iflow diagram for Configuration 3.} \label{fig:billintonCaseE3b}]
      {\scalebox{0.55}{
\begin{tikzpicture}
 \SetUpEdge[lw         = 1pt,
            color      = black,
            labelcolor = white]
  \GraphInit[vstyle=Normal] 
  
  \tikzset{VertexStyle/.append  style={fill,thick}}
  \tikzset{mynode/.style=VertexStyle}

  \node[] (s1) at (0,-1.5) {};
  \node[] (s2) at (1.5,-1.5) {};
  \draw[ultra thick,directed1,blue!50] (s1)--(s2) node [midway,above,black] {iflow};
  \node[] (s3) at (0,-2.5) {};
  \node[] (s4) at (1.5,-2.5) {};
  \draw[ultra thick,-triangle 45,black!50] (s3)--(s4) node [midway,above,black] {islack};
  \node[] (s5) at (0,-3.5) {};
  \node[] (s6) at (1.5,-3.5) {};
  \draw[ultra thick,-open triangle 45,black!50] (s5)--(s6) node [midway,above,black] {$\theta_i$};

  \SetGraphUnit{3}
  \tikzset{VertexStyle/.append  style={fill,thick}}
  \Vertex[x=0 ,y=0]{0}
  \Vertex[x=3 ,y=0]{1}
  \Vertex[x=6 ,y=0]{2}
  \Vertex[x=9 ,y=0]{3}
  \node[mynode,label={DG}] (4) at (12,0) {4};
  \Vertex[x=3 ,y=-3]{5}
  \Vertex[x=6 ,y=-3]{6}
  \Vertex[x=9 ,y=-3]{7}
  \tikzset{EdgeStyle/.style={->}}
  \draw[ultra thick,directed1,red!0!blue!50] (1)--(0) node [midway,above,black] {$0.0$};
  \draw[ultra thick,directed1,red!0!blue!50] (2)--(1) node [midway,above,black] {$0.0$};
  \draw[ultra thick,directed1,red!0!blue!50] (3)--(4) node [midway,above,black] {$0.0$};
  \draw[ultra thick,directed1,red!29!blue!80] (5)--(1) node [midway,right,black] {$0.4$};
  \draw[ultra thick,directed1,red!50!blue!80] (6)--(2) node [midway,right,black] {$1.2$};
  \draw[ultra thick,directed1,red!41!blue!80] (7)--(3) node [midway,right,black] {$0.8$};
  
  \draw [ultra thick,black!50!white,-triangle 45] (1) -- +(-1,1) node [midway,right,black] {$1.2$};
  \draw [ultra thick,black!50!white,-open triangle 45] ($(1) +(1,-1)$) -- (1) node [midway,right,black] {$0.8$};
  \draw [ultra thick,black!50!white,-triangle 45] (2) -- +(-1,1) node [midway,right,black] {$1.6$};
  \draw [ultra thick,black!50!white,-open triangle 45] ($(2) +(1,-1)$) -- (2) node [midway,right,black] {$0.4$};
  \draw [ultra thick,black!50!white,-triangle 45] (3) -- +(-1,1) node [midway,right,black] {$2.0$};
  \draw [ultra thick,black!50!white,-open triangle 45] ($(3) +(1,-1)$) -- (3) node [midway,right,black] {$1.2$};
  \draw [ultra thick,black!50!white,-triangle 45] (5) -- +(-1,1) node [at start,left,black] {$0.0$};
  \draw [ultra thick,black!50!white,-open triangle 45] ($(5) +(1,-1)$) -- (5) node [at start,left,black] {$0.4$};
  \draw [ultra thick,black!50!white,-triangle 45] (6) -- +(-1,1) node [at start,left,black] {$0.0$};
  \draw [ultra thick,black!50!white,-open triangle 45] ($(6) +(1,-1)$) -- (6) node [at start,left,black] {$1.2$};
  \draw [ultra thick,black!50!white,-triangle 45] (7) -- +(-1,1) node [at start,left,black] {$0.0$};
  \draw [ultra thick,black!50!white,-open triangle 45] ($(7) +(1,-1)$) -- (7) node [at start,left,black] {$0.8$};
\end{tikzpicture}
}} 
\end{center}
\caption{Network configurations for which self-interruption $\theta_i$ (hours per year) and power load $l_i$ ($MW$) are given for each node $i$ (left). Iflow diagrams with magnitudes in hours per year (right).}
\label{fig:billintonCaseE1}
\end{figure*}

The distribution networks shown in Figure~\ref{fig:billintonCaseE1}, proposed by \cite{billinton}, showcase the iflows capability of expressing the network's reliability state. 
Different configurations with respect to the location of switches are considered, and the parameters are presented in Table~\ref{tbl:billintonParameters}. 

\begin{table}[ht!]
\setlength{\tabcolsep}{6pt}
\caption{Parameters used in the numerical examples} \begin{center} {
\begin{tabular}{ c c c c c c c c c c } \label{tbl:billintonParameters}
\\ \hline
 &  \multicolumn{9}{c}{Node $i$} \\
 &  & $1$ & $2$ & $3$ & $4$ & $5$ & $6$ & $7$ & $8$ \\ \cline{3-10}
$l_i$ &  & 0.0 & 0.0 & 0.0 & 0.0 & 5.0 & 4.0 & 3.0 & 2.0 \\ 
$\tilde{l}_i$ &  & 14.0 & 9.0 & 5.0 & 2.0 & 5.0 & 4.0 & 3.0 & 2.0 \\ 
$\lambda_i$ &  & 0.2 & 0.1 & 0.3 & 0.2 & 0.2 & 0.6 & 0.4 & 0.2 \\ 
$t_i$ &  & 4.0 & 4.0 & 4.0 & 4.0 & 2.0 & 2.0 & 2.0 & 2.0 \\ 
$\theta_i$ &  & 0.8 & 0.8 & 1.2 & 0.8 & 0.4 & 1.2 & 0.8 & 0.4 \\ 
\hline
\multicolumn{10}{l}{$l_i$ -- node $i$ power load ($MW$).} \\
\multicolumn{10}{l}{$\tilde{l}_i$ -- node $i$ downstream power load ($MW$).} \\
\multicolumn{10}{l}{$\lambda_i$ -- node $i$ failure rate (failures per year).} \\
\multicolumn{10}{l}{$t_i$ -- average interruption time (hours).} \\
\multicolumn{10}{l}{$\theta_i$ -- node $i$ self-interruption (hours per year).} \\
\end{tabular} } 
\end{center} \end{table}

Represented by Figure~\ref{fig:billintonCaseE1a}, Configuration 1 contains only the substation switch, which means that any fault  disconnects all the load points. The iflows under this configuration are shown in Figure~\ref{fig:billintonCaseE1b}, with a maximum iflow of $4.8$ $hours/year$. This figure  illustrates the balance among iflows, islacks, and self-interruptions, as predicted by Eq.~\eqref{eq:varF}.
Represented by Figure~\ref{fig:billintonCaseE2a}, Configuration 2 contains four lateral switches. Any fault on Nodes $5$, $6$, $7$, or $8$ is contained by their corresponding switch. Figure~\ref{fig:billintonCaseE2b} shows a distinct improvement  in the overall values of the iflows; compared to Configuration 1, the maximum iflow is reduced by half, to $2.4$ $hours/year$.
Represented by Figure~\ref{fig:billintonCaseE3a}, Configuration 3 considers a distributed generation power source located at Node 4 with a capacity of $8.0$ $MW$. The values of iflows are obtained likewise Configurations 1 and 2, as shown in Figure~\ref{fig:billintonCaseE3b}. 

\tikzstyle arrowstyle=[scale=1.5]
\tikzstyle directed1=[postaction={decorate,decoration={markings,
mark=at position .5 with {\arrow[arrowstyle]{stealth}}}}]
\tikzstyle directed2=[postaction={decorate,decoration={markings,
mark=at position 1 with {\arrow[arrowstyle]{stealth}}}}]

The ENS can be computed using Eq.~\eqref{eq:ENS2}, which for the network topology of Configurations 1 and 2 leads to
\begin{align} 
	\displaystyle ENS =& \ (\tilde{l}_1-\tilde{l}_2)f_{12} + (\tilde{l}_1-\tilde{l}_5)f_{15} + (\tilde{l}_2-\tilde{l}_3)f_{23} + (\tilde{l}_2-\tilde{l}_6)f_{26} \nonumber \\
	&+ (\tilde{l}_3-\tilde{l}_4)f_{34} + (\tilde{l}_3-\tilde{l}_7)f_{37} + (\tilde{l}_4-\tilde{l}_8)f_{48} + 
	\tilde{l}_1\theta_1 + \tilde{l}_2\theta_2 \nonumber \\
	&\tilde{l}_3\theta_3 + \tilde{l}_4\theta_4 + \tilde{l}_5\theta_5 + \tilde{l}_6\theta_6
	+ \tilde{l}_7\theta_7 + \tilde{l}_8\theta_8 
	\label{eq:example}
\end{align}	
With respect to Configuration 3, given the presence of an additional power source, a more attentive analysis is called for. 
It should be noticed that Nodes $2$ and $6$ are insulated from interruptions at Nodes $1$ and $5$ by opening the switch in arc $(1,2)$ and closing the switch in arc $(2,3)$. This is addressed by using the appropriate power load coefficients in Eq.~\eqref{eq:ENS2}, which for Configuration 3 leads to

\begin{align} 
	\displaystyle ENS =& \ (\tilde{l}_1-\tilde{l}_2)f_{12} + (\tilde{l}_1-\tilde{l}_5-\tilde{l}_2)f_{15} + (\tilde{l}_2-\tilde{l}_6)f_{26} + (\tilde{l}_3-\tilde{l}_7)f_{37} + \nonumber \\
	&+ (\tilde{l}_1-\tilde{l}_2)\theta_1 + \tilde{l}_2\theta_2 + \tilde{l}_3\theta_3 +  \tilde{l}_5\theta_5 + \tilde{l}_6\theta_6	+ \tilde{l}_7\theta_7
	\label{eq:example2}
\end{align}	
By substituting the values of the iflows from Configurations 1 and 2 in Eq.~\eqref{eq:example} and from Configuration 3 in Eq.~\eqref{eq:example2}, an ENS of $84.0$ $MWh/year$, $54.8$ $MWh/year$, and $18.4$ $MWh/year$, respectively, are obtained. It should be noticed that these results match with the results obtained by \cite{billinton}.

\section{Case study on the switch allocation problem} \label{sec:Optimization}

The switch allocation problem (SAP) aims towards the best  locations of switches on a distribution network considering  cost-reliability trade-offs. The potential benefits of switch allocation include a reduction in the average duration of failures, an improvement in the quality of the power supply, and the avoidance of fines related to the violation of reliability standards.
Here, we consider the allocation of switches to minimize the ENS. 

Many heuristics have been proposed to solve the SAP. \cite{levitin1994} introduced this optimization problem and proposed a genetic algorithm to allocate sectionalizers in a radial distribution network. 
Several other researchers followed the ideas of \cite{levitin1994} in their proposals of metaheuristics for the SAP, such as simulated annealing \cite{billinton2}, iterated sample construction with path relinking \cite{Benavides2013}, memetic algorithm \cite{assis2014}, and bee colony algorithm 
\cite{AmanEtal2016}.
\cite{JahromiEtal2012} proposed a mixed-integer linear model with an explicit enumeration of the locations of switches, resulting in an exponential number of variables and constraints.
\cite{FarajollahiFotuhiSafdarian2019} extended the previous model also to consider the allocation of fault indicators. The authors conclude that the allocation of switches is more  effective cost-wise.

As shown in the following section, by
using the iflows, a polynomial-size mixed-integer programming model with a better scaling capability is attained.

\subsection{Mixed-integer programming model} \label{sec:model}

A fixed number of $N$ switches is considered in the optimization problem to minimize the ENS.
A switch is on arc $(i,j)$ if and only if $x_{ij} = 1$. Variable $f_{ij}$ gives the iflow (Eq.~\ref{eq:flow}) on arc $(i,j)$. Parameters $\tilde{l}$, $\theta_i$, and $E_{lb}$ are described by Eqs.~\eqref{eq:tildepi}, \eqref{eq:selftime}, and \eqref{eq:Eminmax}, respectively. A sufficiently large constant $M_i$ is defined for each node $i$. 
\begin{align}
	& \mbox{(SAP)} & \notag \\
	& MIN \quad \sum_{(i,j) \in A}{(\tilde{l}_i-\tilde{l}_j)f_{ij}}+E_{lb} \label{eq:minSAPa}  \\
	& \mbox{s.t.} & \notag \\
	& \mbox{\textit{\small (number of switches)}} & \notag \\
 	& \sum_{(i,j) \in A}{x_{ij}} \leqslant N & \label{eq:minSAPb}  \\
	& \mbox{\textit{\small (iflow node balance)}} & \notag \\
	& \displaystyle F_j + f_{ij} = \theta_j + \sum_{(j,k) \in A}{f_{jk}} & (i,j) \in A \label{eq:minSAPc} \\
    & \mbox{\textit{\small (islack coupling with switch allocation)}} & \notag \\
    & \displaystyle F_j \leqslant M_jx_{ij} & (i,j) \in A \label{eq:minSAPd} \\
	& \mbox{\textit{\small (variables bounds and integrality)}} & \notag \\
	& f_{ij}, F_{j} \geqslant 0, \ \ x_{ij} \in \left\{0,1\right\} & (i,j) \in A \label{eq:domain1} 
\end{align}

The objective function \eqref{eq:minSAPa} represents the solution ENS. The number of switches is constrained by \eqref{eq:minSAPb}. The iflow node balance is expressed in \eqref{eq:minSAPc}. Constraints \eqref{eq:minSAPd} couple the decision of allocating a switch in an arc with the value of the corresponding islack. If an arc $(i,j)$ does not contain a switch ($x_{ij} = 0$), the islack $F_j$ is forced to zero, while the iflow assumes the value predicted by the node balance. Conversely, if arc $(i,j)$ does contain a switch, the value of $F_j$ must be allowed to assume a sufficiently large value to absorb all downstream interruptions, thus allowing $f_{ij}$ to be zero. In the worst case, the amount of downstream interruptions that the islack should absorb is equal to the maximum iflow of arc $(i,j)$ (Eq.~\eqref{eq:maxflow}). Thus, $\displaystyle M_i = \sum_{j \in V_i}{\theta_j}$ was set in the computational experiments described in the next section.

\subsection{Computational experiments} \label{sec:Tests}

The SAP is solved for a benchmark of radially-operated distribution networks\footnote{Data available at the address (last accessed in May 2021): \\ http://www.dejazzer.com/reds.html}, the attributes of which are described in Table~\ref{tbl:networkAttributes} \cite{redes2013}. 
Solutions are obtained with Gurobi 8.1 under a time limit of ten minutes, on an Intel i7 3930k with 16 GB of RAM, and Ubuntu 18.04.

\begin{table}[ht!]
\caption{Benchmark networks attributes} 
\begin{center} {
\begin{tabular}{ c c c c c c } \label{tbl:networkAttributes}
\\
Network & $|V|$ & $|A|$ & $\tilde{l}_0$ & $E_{lb}$ & $E_{ub}$ \\ \hline
R3 & 34 & 33 & 3,708.27 & 2,069.97 & 11,135.23 \\ 
R4 & 95 & 94 & 28,342.96 & 2,340.32 & 4,242.33 \\ 
R5 & 144 & 143 & 18,315.82 & 3,747.42 & 14,110.97 \\ 
R6 & 205 & 204 & 27,571.37 & 1,437.63 & 6,932.57 \\ 
R7 & 881 & 880 & 124,920.01 & 266,293.63 & 1,518,308.94 \\ 
\hline 
\multicolumn{6}{l}{$\tilde{l}_0$ -- total power load ($kW$) Eq.~\eqref{eq:tildepi}.} \\
\multicolumn{6}{l}{$E_{lb}$ -- ENS lower bound ($kWh$/year) Eq.~\eqref{eq:Eminmax}.} \\
\multicolumn{6}{l}{$E_{ub}$ -- ENS upper bound ($kWh$/year) Eq.~\eqref{eq:Eminmax}.}  \\
\end{tabular} }
\end{center} \end{table}

The maximum number of switches in a solution is set to
$N = \lfloor P \cdot |A| \rfloor$. Five networks and four values of $P$ result in a total of $20$ instances. Table~\ref{tbl:results} gives the objective function (ENS) of the best feasible solutions ($UB$), the optimality gap ($Gap$), and the execution times ($CPU$).

The SAP model obtains optimal solutions for 11 instances with up to 205 nodes, taking less than one second of computation. The model was unable to prove optimality for nine instances, with an average gap of $2.01\%$ and a worst gap of $8.21\%$. 
The results indicate that the model is sensitive to the size of the instance, as well as to the number of switches allowed. The solution gap decreases as the number of switches increases, with only one exception (network R5, $P = 40\%$).

\begin{table}[ht!]
\caption{Computational results summary}
\begin{center} {
\begin{tabular}{ c c c c p{0.1cm} c c c } \label{tbl:results}
\\ 
 \cline{1-8}
 & \multicolumn{3}{c}{$P = 20\%$} &  & \multicolumn{3}{c}{$P = 40\%$} \\ \cline{2-4} \cline{6-8} 
 Network &  $ENS$ & $Gap$ & $CPU$ & & $ENS$ & $Gap$ & $CPU$ \\ \hline
R3 &  \textbf{2715.24} & \textbf{opt} & $<$ 1 &  & \textbf{2269.21} & \textbf{opt} & $<$ 1  \\ 
R4 &  \textbf{2504.72} & \textbf{opt} & $<$ 1 &  & \textbf{2361.50} & \textbf{opt} & $<$ 1   \\ 
R5 &  \textbf{4801.43} & \textbf{opt} & $<$ 1 &  & 3928.03 & 2.47 & 600 \\ 
R6 & 1661.86 & 2.24 & 600 &  & 1457.86 & 1.11 & 600 \\ 
R7 &  307668.14 & 8.21 & 600 &  & 274710.83 & 2.92 & 600 \\ 
 
  & \multicolumn{3}{c}{$P = 60\%$} &  & \multicolumn{3}{c}{$P = 80\%$} \\ \cline{2-4} \cline{6-8} 
 Network  & $ENS$ & $Gap$ & $CPU$ & & $ENS$ & $Gap$ & $CPU$ \\ \hline
R3 &  \textbf{2144.88}  & \textbf{opt} & $<$ 1 &  & \textbf{2089.06} & \textbf{opt} & $<$ 1  \\ 
R4 &  \textbf{2340.32} & \textbf{opt} & $<$ 1 &  & \textbf{2340.32} & \textbf{opt} & $<$ 1 \\ 
R5 &   3774.88 & 0.31 & 600 &  & \textbf{3747.42} & \textbf{opt} & $<$ 1 \\ 
R6 &  1439.19 & 0.05 & 600 &  & \textbf{1437.63} & \textbf{opt} & $<$ 1 \\ 
R7 &  268135.66 & 0.69 & 600 &  & 266548.57 & 0.09 & 600 \\ \hline
\multicolumn{8}{l}{$P$ -- percentage of switches with respect to the number of arcs.} \\
\multicolumn{8}{l}{ENS -- energy not supplied ($kWh$/year) of the best feasible solution.} \\
\multicolumn{8}{l}{$Gap = 100\cdot(UB-LB)/UB$.} \\
\multicolumn{8}{l}{$CPU$ -- computational time in seconds.} \\
\end{tabular} } 
\end{center} \end{table}

\begin{figure}[hbtp]
	\begin{center}
	      \includegraphics[width=0.8\textwidth]{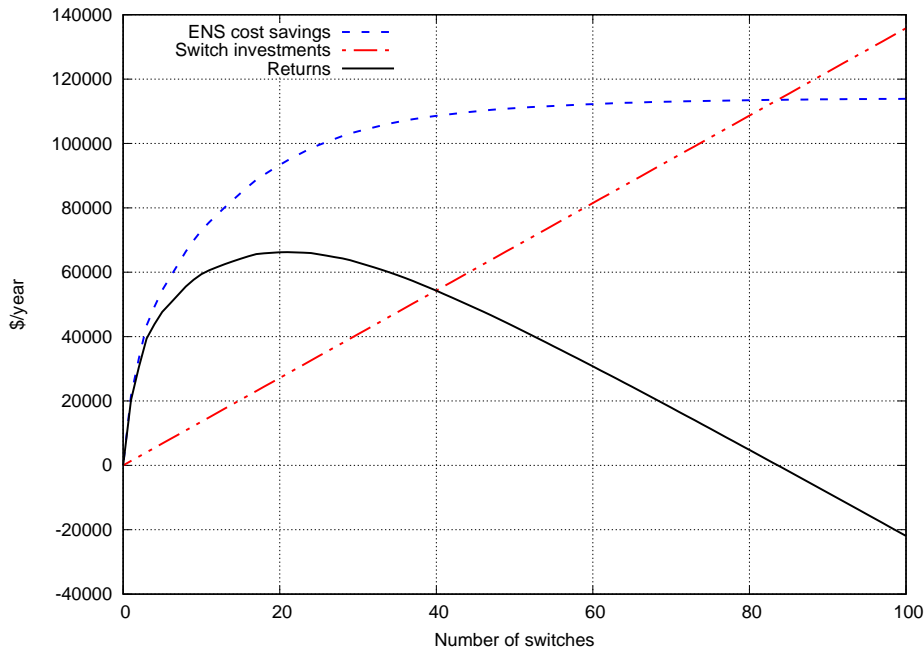}
	\end{center}
\caption{Cost and returns pertaining the switch allocation in Network R5.}
\label{fig:R5example}
\end{figure}

Now consider a greenfield scenario, in which a planner must decide on economic grounds the best number and locations of switches on an empty network. The annual investment to acquire and maintain a sectionalizer and the cost of interrupted energy are estimated as $C_s =$ US\$$1,358.00/year$ and $C_e =$ US\$$1.53/kWh$, respectively \cite{brown2008}. Figure~\ref{fig:R5example} shows the ENS cost savings $(C_e \cdot (E_{ub} - E_N))$ and the switch investment $(C_s \cdot N)$ for solutions of Network R5 containing $N = [0,\ldots,100]$ switches. By subtracting the switch investment from the ENS cost savings, the annual returns enabled by the allocation of switches are obtained. For Network R5, solutions with up to $83$ switches have positive returns, which means the reliability investment pays for itself in these cases. The best solution is to allocate $21$ switches, giving a maximum return of US\$$66,247.41/year$.

These numerical experiments support the model's strength in enabling exact solutions to large instances with more than 800 nodes. Previous approaches either rely on heuristic methods or solve instances of modest sizes -- smaller than the problem with 144 vertices addressed  by \cite{FarajollahiFotuhiSafdarian2019}. The strength of the model lies in the relationship between interruptions and network flow, which the concept of iflows renders straightforward.

\vspace{-3.0mm}
\section{Final Remarks} \label{sec:Conclusions}

The paper advocates the concept of iflows, which supports a new reliability evaluation framework for energy networks. These flows can be computed efficiently in linear time, and they readily provide information about reliability indices. This aspect may be crucial for modern energy networks such as smart grids, demanding frequent and expeditious reliability evaluations.

The iflows embody essential reliability states allowing an analytic evaluation through a series of flow balance computations.
The framework was employed in a mixed-integer linear model for a switch allocation problem. 
The results of case studies using a benchmark of distribution networks show 
that high-quality solutions, most of them optimal, can be  obtained within
short computational times. This result sets a new standard for solving the switch allocation problem, opening  paths for research on polyhedral aspects of the model, and its facet-defining inequalities. Moreover, reducing the reliability evaluation as a network flow problem suggests the design of customized algorithms for the switch allocation problem.

The use of iflows in energy
networks should be developed in future studies to encompass other reliability optimization problems, which include contingency planning, maintenance scheduling, self healing, fault indicator allocation, and the effects of switching interruptions, variable topology of meshed networks, islanding operation, and component failure. 
Also, a promising line of research is to extend the proposed framework to other service networks such as sensor networks, and smart metering.

\section*{Acknowledgments}
This work was supported by the Brazilian research agency CNPq (proc. 435520/2018-0, 312647/2017-4) and Fapesp (proc. 2015/11937-9, 2016/08645-9).

\bibliographystyle{alpha}
\bibliography{bibliografia}

\end{document}